# Observation of floating surface state in obstructed atomic insulator candidate NiP$_2$

Xiang-Rui Liu[1*], Ming-Yuan Zhu[1*], Yuanwen Feng[2,3*], Meng Zeng[1], Xiao-Ming Ma[1], Yu-Jie Hao[1], Yue Dai[1], Rong-Hao Luo[1], Kohei Yamagami[4], Yi Liu[5], Shengtao Cui[5], Zhe Sun[5], Jia-Yu Liu[6], Zhengtai Liu[7], Mao Ye[7], Dawei Shen[5,8], Bing Li[2,3†], Chang Liu[1†]

[1]*Department of Physics and Shenzhen Institute for Quantum Science and Engineering (SIQSE), Southern University of Science and Technology (SUSTech), Shenzhen, Guangdong 518055, China.*
[2]*Shenyang National Laboratory for Materials Science, Institute of Metal Research, Chinese Academy of Sciences, 72 Wenhua Road, Shenyang, Liaoning 110016, China.*
[3]*School of Materials Science and Engineering, University of Science and Technology of China, 72 Wenhua Road, Shenyang, Liaoning 110016, China.*
[4]*Japan Synchrotron Radiation Research Institute (JASRI), Sayo, Hyogo 679-5198, Japan.*
[5]*National Synchrotron Radiation Laboratory, University of Science and Technology of China, Hefei, Anhui 230029, China*
[6]*National Key Laboratory of Materials for Integrated Circuits, Shanghai Institute of Microsystem and Information Technology, Chinese Academy of Sciences, Shanghai 200050, China*
[7]*Shanghai Synchrotron Radiation Facility, Shanghai Advanced Research Institute, Chinese Academy of Sciences, Shanghai 201210, China*
[8]*School of Nuclear Science and Technology, University of Science and Technology of China, Hefei, Anhui 230026, China*

*These authors contribute equally to this work.
†Corresponding authors.

# Abstract

Obstructed atomic insulator is recently proposed as an unconventional material, in which electric charge centers localized at sites away from the atoms. A half-filling surface state would emerge at specific interfaces cutting through these charge centers and avoid intersecting any atoms. In this article, we utilized angle-resolved photoemission spectroscopy and density functional theory calculations to study one of the obstructed atomic insulator candidates, $NiP_2$. A floating surface state with large effective mass that is isolated from all bulk states is resolved on the (100) cleavage plane, distinct from previously reported surface states in obstructed atomic insulators that are merged into bulk bands. Density functional theory calculation results elucidate that this floating surface state is originated from the obstructed Wannier charge centers, albeit underwent surface reconstruction that splits the half-filled obstructed surface state. Our findings not only shed lights on the spectroscopy study of obstructed atomic insulators and obstructed surface states, but also provide possible route for development of new catalysts.

In the past decades, various topological phases were theoretically proposed and experimentally realized in condensed matter systems, characterized by some non-zero topological invariants defined in the reciprocal space. These topological phases include but not limit to topological insulators [1, 2], quantum anomalous Hall insulators [3] and topological semimetals [4]. Recently, systematic topological classification based on different theoretical methodologies and high-throughput calculations were performed to filter out all possible topological materials from the materials database [5-8]. However, even in materials that are classified to be "topologically-trivial" according to *k*-space invariants, part of them still differs from others in a topological sense. These materials are characterized by non-zero parameters defined in the *real space* [9, 10], rendering characteristic electronic structures that cannot be adiabatically evolve to the equivalency of those formed by separated atoms.

One of such unconventional types of materials has been defined as "obstructed atomic insulators" (OAIs), attracting broad scientific attention [9-15]. Distinct from the ordinary atomic insulators, there are Wannier charge centers (sites at which the electrical charge spatially localizes and exponentially decays away) located at some empty sites in the crystal of an OAI, which are characterized by non-zero real space invariants [9, 10]. For specific interfaces that only cut through such obstructed Wannier charge centers (OWCCs), fractionalization of charge would occur at the interface, according to the filling anomaly principle [16-19]. Ideally, this would result in the emergence of low-bandwidth surface electronic states that floats between the bulk valence and conduction bands and cuts through the Fermi level regardless of electronic occupation. It is these "obstructed surface states (OSSs)" that contribute predominantly to the catalytic activity of OAIs. High-throughput first-principles calculations have screened out hundreds of OAI candidates from the current material database [10, 11, 15]. These candidates can be identified as various functional materials, including electrides [20-23], thermoelectric materials, hydrogen evolution reaction (HER) electrocatalysts [10, 24], etc. The diversity of OAI candidates paves the way for investigation and application of them in the future.

Currently, the theoretical description of OAIs has been well developed, while the experimental study on them is still rare. A recent study verified the catalytic active sites in HER of several OAI candidates [24]; another work observed the in-gap soliton state at the domain boundary on Si 2×1 (111) surface, which can be manipulated by electric pulse discharged from the tip [25]. These results support the theoretical predictions of OAIs, elucidating the relation between novel phenomena and localized OWCCs, and revealing the application potential of OAIs on catalysis. Noticeably, spectroscopic experiments reveal that the OSSs in OAI candidates are subject to significant modifications in realistic surfaces that undergo structural reconstructions. Such reconstructions would remedy the fractionalization of charges, causing an energy split of the OSSs to the level that the two branches locate at both sides of the Fermi level. In the case of $SrIn_2P_2$, the upper branch of the modified OSSs on the $\sqrt{3} \times 1$ (0001) plane show negative differential conductance following a strong $dI/dV$ peak above $E_F$ [26]. On the other hand, the lower branch in both $SrIn_2P_2$ and Si (111) are seen to not only locate mostly below $E_F$ but also connects to the bulk valence bands [25, 26]. The entanglement between the bulk states and the SSs hampers the distinguishability and application of these electronic states. Thus, finding OWCC-originated SSs that are separated from bulk states in OAI candidates is demanded to complement the knowledge about this type of unconventional materials and instruct the development of new catalysts.

In this article, combined with angle-resolved photoemission spectroscopy (ARPES) and density functional theory (DFT) calculations, the electronic structure of monoclinic $NiP_2$ (denoted as *m*-$NiP_2$) crystals were systematically studied. The dispersion of bulk bands and a global flat band in the three-dimensional Brillouin zone (BZ) are observed. Importantly, a floating SS (fSS) with large effective mass that is separated from other bands on the (100) surface is resolved near the surface Brillouin zone (sBZ) center, and it evolves to be even flatter at the sBZ boundary. The temperature dependence of band structures on the (100) surface of *m*-NiP$_2$ is also studied, which is found to be in line with the temperature evolution of the electrical conductivity. The DFT calculation results reproduce the dispersion of bulk states and the fSS on the (100) plane. Our findings elucidate the relation between the fSS and OWCCs in *m*-NiP$_2$, highlighting the importance of fully separated SSs on OAI candidates, and providing an explanation of the termination-dependent catalytic activity from the electronic structure point of view.

As one of the OAI candidates [10], *m*-NiP$_2$ crystallizes in space group C2/*c*. The nickel and phosphorous atoms form a single atomic layer within the *bc* plane, and the neighboring layers are coupled via P-P chains along the *a* axis (Fig. 1 (a, b)) [27]. The theoretically predicted OWCCs locate at Wyckoff positions 4*d* (between two neighboring phosphorous atoms within a single atomic layer) and 4*e* (between two neighboring phosphorous atoms along the P-P chain), while Ni and P atoms locate at Wyckoff positions 4*c* and 8*f*, as shown in Fig. 1. Both the (100) and (11-1) cleavage plane cut through the OWCCs and avoid intersecting any atoms. (Here, the Miller indexes of these planes are defined in the conventional cell, rather than the primitive unit cell.) According to the theoretical prediction, fractional charge and OSS would emerge once the crystal is terminated at these planes. Moreover, as seen in Fig. 1(a), the cleavage along the (100) plane not only cut through the OWCCs at 4*e* (magenta spheres), but also expose the OWCCs at 4*d* (orange spheres) to the vacuum. In principle, the fully occupied spatially-localized charge centers (OWCCs at 4*d*) could also serve as sites for potential ligand adsorption and charge transfer, and contribute to the interface catalysis. In a recent study, the (100) surface of *m*-NiP$_2$ shows much higher catalytic activity than other terminations in HER [28], which supports this hypothesis and implies the essential role of OWCCs in interface catalysis.

We start from the data collected on samples cleaved along the (100) surface measured using soft X-ray photons at 80 K. The normal vector of the in-plane Brillouin zone is determined to be the [1-10] direction, in units of primitive unit vectors in the reciprocal space. For details about the three-dimensional Brillouin zone, in-plane Brillouin zone and sBZ, please read Section I of the Supplementary Material. The ARPES $E_B$-$k_x$ band dispersion near the Fermi level, which cuts through different positions in the three-dimensional Brillouin zone, are shown in Fig. 2. It's easy to find one relatively flat, wave-like floating band that tops at $E_B \sim 0.2$ eV (denoted as the floating surface state, fSS), separating from other bands in energy. It disperses as a hole-like pocket at the $\bar{\Gamma}$ point, with a global band width of only about 300 meV. This band shows almost identical dispersion along $k_x$ in several sBZs (Fig. 2(a)), as well as along $k_z$ at several out-of-plane Brillouin zones (Figs. 2(c, d)), signifying its surface origin. There is also an electron-like band at the $\bar{\Gamma}$ point whose bottom locates at $E_B \sim 1.8$ eV (denoted as e1), which appears to be the same at different $k_z$s. This is also a surface band. On the other hand, two hole bands whose band tops locate at $E_B \sim 0.8$ and 1.3 eV (denoted as h1 and h2, respectively) are observed at the $\bar{\Gamma}$ point. These pockets show strong dispersion across several sBZs and different $k_z$s, implying their bulk origin. Besides these dispersive bands, one could also notice a nearly flat band that locates at $E_B \sim 2.0$ eV at different $k_z$s. As discussed later, this band is a bulk global flat band that appears in the whole three-dimensional Brillouin zone.

The photon energy-dependent map was performed for direct visualization of the out-of-plane dispersion (Fig. 2 (c1, c2)). The out-of-plane momentum is calculated via $k_z = \sqrt{2m/\hbar^2 (E_k \cos\theta^2 + V_0)}$, where the inner potential is determined to be $V_0 = 12$ eV. The $E_B$-$k_z$ dispersion collected at the $\bar{\Gamma}$ point of the first and the second sBZ (corresponds to $k_x = 0$ and 1.119 Å$^{-1}$, respectively) verifies that both the fSS and the bulk flat band show no $k_z$ dispersion over the entire three-dimensional Brillouin zone. On the other hand, the h1 band shows well-defined periodic $k_z$ dispersion within the first sBZ. It is identified as a single hole pocket near the Γ point and evolves to a "M"-shaped band near the Z point. The periodicity in the second sBZ reveals a π phase shift compared with that in the first sBZ. Such feature coincides with the in-plane Brillouin zone evolution behavior at different $k_z$s (Section I in the Supplementary Material), suggesting the bulk nature of the h1 band. Upon a global energy shift of -0.28 eV which accounts for possible charged defects, the DFT calculation results agree well with the observed dispersion of the h1 band and the flat band, further confirming their bulk nature. The fSS and the e1 pocket are absent in the bulk-band calculation results, but are present in the DFT calculation based on an 8-unit-cell slab model, which further elucidates their surface nature.

To study the band dispersion along the in-plane directions, $k_x$-$k_y$ maps are performed at selected photon energies. The constant energy contours (CECs) at $E_B = 0.4$ eV collected with 116 eV incident photons reveal a spot-like feature between two adjacent fSS pockets along the $k_y$ direction (indicated by the black arrow in Fig. 3 (a)). In the $E_B$-$k_y$ dispersion that crosses the $\bar{\Gamma}$ point at 116 and 83 eV (correspond to $k_z = 5.694$ Å$^{-1}$, close to the Z point, and 4.875 Å$^{-1}$, close to the Γ point, respectively), it's easy to identify it as another hole pocket (Fig. 3 (b, c)). This pocket is not clearly resolved in the data collected within the soft X-ray energy range, possibly due to the reduced photoemission cross-section at higher photon energies (Section III in the Supplementary Material). Both hole pockets of the fSS show no $k_z$-dispersion from 24 to 120 eV (corresponds to $k_z = 2.892$ to 5.785 Å$^{-1}$, over an entire three-dimensional Brillouin zone), indicating that it should be a SS (Section II in the Supplementary Material). The effective masses of the two pockets of fSS are estimated to be -4.85 $m_e$ and -1.45 $m_e$, respectively (Section IV in the Supplementary Material).

Apart from the previously mentioned spot-like feature, a ribbon-like, faint trace that connects the fSS hole pockets between neighboring sBZs along the $k_x$ direction also presents in the constant energy contours, as indicated by the white arrow and white dashed line in Fig. 3 (a). Thus, we examine the dispersion of the fSS near the boundary of the sBZ. From Figs. 3(d)-(g), it is easy to find that the fSS evolves to be an almost flat band at the M-X-M cuts along $k_y$ (Figs. 3 (d, e)), and a single hole pocket centered at the Y point at the M-Y-M cuts along $k_x$ (Figs. 3 (f, g)). All of the energy distribution curves (EDCs) tracked from the spectrum collected at 116 eV reveal a peak feature at a constant binding energy $E_B = 0.55$ eV, which further confirm its flat band nature. As for the case of 83 eV, the spectral weight near $k_y = 0$ is weak due to the matrix element effect [29], and the flatness of such band in other momentum range is reserved. The spatially-localized OWCCs at the interface would yield surface bands that disperse with low group velocity, i.e., these bands possess relatively large effective mass. Thus, the large effective mass of the hole pocket at the center of the sBZ also reflects the relation between the fSS and OWCCs. Our DFT calculation results on an 8-unit-cell slab model reveal a floating SS within the bulk band gap, well reproducing the observed dispersion of the fSS along both the $k_x$ and $k_y$ directions (Fig. 3 (i, j)). The fSS found in our ARPES measurements and DFT calculations originates from the OWCCs, albeit underwent an energy splitting possibly due to surface structural reconstruction, similar to the case in SrIn$_2$P$_2$ [26]. Important differences between the OSSs in the two materials are that (1) the lower branch of the reconstructed surface band merges to the bulk states in the case of SrIn$_2$P$_2$, while remaining separated from the bulk bands in the case of *m*-NiP$_2$; (2) the OSS in *m*-NiP$_2$ exhibit much smaller bandwidth than that in SrIn$_2$P$_2$, and (3) the OSS in *m*-NiP$_2$ locates in general closer to $E_F$ than that in SrIn$_2$P$_2$. These features make the OSS-originated fSS in *m*-NiP$_2$ more relevant to its transport properties and thus more related to the enhancement of the catalytic activity in the (100) surface.

To investigate the correspondence between the band structure and the electrical transport, we studied the temperature-dependence of the bands on the (100) plane of *m*-NiP$_2$ from T = 8 to 150 K, with an incident photon energy of 42 eV. The EDCs collected at the $\bar{\Gamma}$ point at different temperatures were shown in Fig. 4 (a). Here, we set the band position at 20 K as a reference, and the band shift of the fSS and the e1 pocket are obtained by fitting the EDC peaks at different temperatures. Both the fSS and the e1 pocket are found to shift upward upon heating, and recover to previous energy upon cooling. These two pockets shift with temperature at the same pace, revealing a rigid band shifting behavior. This shift is relatively acute below 100 K and becomes moderate above 100 K. In our transport measurement, the electrical conductance of $m$-NiP$_2$ increases dramatically with increasing temperature from 50 to 100 K, and the slope decreases within 100 to 150 K (Fig. 4 (b)). The agreement between the temperature evolution of the conductance and that of the band shift suggests a close relationship between the transport behavior and the electronic structure, especially the fSS, in the $m$-NiP$_2$ (100) surface.

To further discuss the relation between the surface electronic structure and the termination-dependent catalytic activity, the band structure of $m$-NiP$_2$ that cleaved along the (11-1) plane was also studied (Fig. S7, S8). Although faint traces of possible electronic states are found to cross $E_F$ after second-order curvature analysis [30], it is not certain whether these are the SSs of the $m$-NiP$_2$ (11-1) cleavage plane. Nonetheless, the divergent surface electronic structures on different cleavage planes of $m$-NiP$_2$ are likely responsible for the previously reported termination-dependent catalytic activity.

According to theoretical prediction, both the (100) and (11-1) planes would cut through the OWCCs in $m$-NiP$_2$ without intersecting any atoms, yielding fractional charges and metallic SSs at the interface. However, in realistic situations, the fractional charges localized at the interface may not be stable, and are subject to surface reconstruction. The spontaneous surface reconstruction upon cleavage would change the charge distribution on the surface, reshaping or even eliminating the theoretically predicted half-occupied OSSs. In previously reported ARPES studies on OAI candidates, the OSSs evolved to a fully-occupied branch and a fully-unoccupied branch to lower the total energy of the surface [25, 26]. The fSS on the (100) surface of $m$-NiP$_2$ should also be ascribed to this situation. In Fig. 5, we investigate the electrical charge distribution of the fSS, by performing DFT calculation an 8-unit-cell slab model, and mapping the charge distribution of the bands within $E_B$ = 0 to 0.5 eV (where the occupied fSS locates in energy) to the real space (Fig. 5(b)). From Fig. 5(a), one realized that the original

OWCC-induced OSS indeed split into two branches, probably via surface reconstruction. Both branches float within the bulk band gap and isolate from all the bulk states. From Fig. 5(b), we found that the electrical charges (yellow bubbles) mainly locate either around the Ni atoms or just above the P atoms at the interface, which provides solid evidence that the OWCCs localized at 4$e$ are mainly responsible for the observed fSS near the Fermi level. Meanwhile, the observed temperature dependence of the fSS suggests that the lower branch of the fSS would get closer to the Fermi energy at higher temperatures. In principle, these features benefit for the charge transfer in room-temperature catalytic processes. On the other hand, considering the atomic sites in $m$-NiP$_2$, the surface structure of the as-cleaved (11-1) plane is more complex than that of the (100) plane. Thus, the surface relaxation upon cleavage along the (11-1) plane would modify the atomic sites and charge distribution of several topmost atomic layers to a greater extent than that in the (100) plane, which may even eliminate the theoretically predicted OSSs. This distinction can explain the emergence of the fSS near the Fermi energy on the (100) surface rather than the (11-1) surface of $m$-NiP$_2$. Moreover, the (100) surface not only cuts through OWCCs at 4$e$, but also expose the OWCCs at 4$d$ to the interface. Those charge centers can also serve as sites for charge transfer and ligand adsorption, and contribute to the surface catalysis.

In conclusion, we combine ARPES and DFT calculations to study the electronic structure of $m$-NiP$_2$. A fSS that is separated from the bulk bands in energy is observed on the (100) plane, which features a low global bandwidth of ~0.3 eV, and locates only ~0.3 eV below $E_F$ at 20 K. At elevated temperatures, the fSS is found to evolve even closer to $E_F$, which is consistent with electrical conductivity measurements and indicates an important role the fSS plays in the transport process in devices that are functional at room temperature. Our DFT calculation results reproduce the dispersion of both the bulk bands and the fSS, and point out that the observed fSS is the lower, occupied branch of the energy-split SSs that originated from the OWCC-induced OSS. Our findings confirm an isolated, relatively flat SS in one of the OAI candidates, give an explanation on the catalytic performance of *m*-$NiP_2$ from the surface electronic structure point of view, and offer a novel idea for the development of new catalysts.

**Acknowledgements**

This work was supported by the National Key R&D Program of China (Grant Nos. 2022YFA1403700, 2020YFA0308900, 2021YFB3501201 and 2023YFA1406304), the National Natural Science Foundation of China (NSFC) (Nos. 11934007, 12074161, 12204221), NSFC Guangdong (No. 2016A030313650, 2022A1515012283), the Key-Area Research and Development Program of Guangdong Province (2019B010931001), and the Guangdong Innovative and Entrepreneurial Research Team Program (Nos. 2016ZT06D348, 2017ZT07C062). The ARPES experiments were performed at BL03U of Shanghai Synchrotron Radiation Facility under the approval of the Proposal Assessing Committee of SiP.ME$^2$ platform project (Proposal No. 11227902) supported by NSFC. The DFT calculations were performed at Center for Computational Science and Engineering of Southern University of Science and Technology. D.S. acknowledges support from NSFC (No. U2032208). C.L. acknowledges support from the Highlight Project (No. PHYS-HL-2020-1) of the College of Science, SUSTech.

**Methods**

*Sample growth*

Single crystals of *m*-NiP$_2$ were grown using the Sn-flux method [27]. The Ni powder (99%), P powder (Alfa Aesar, 99%) and Sn shots (Aladdin, 99.5%) was mixed with a molar ratio Ni : P : Sn = 1 : 2 : 20 in an alumina crucible, and then flame sealed in quartz ampoule under Argon protection. The sealed ampoule is set in a box furnace. The furnace is heated from room temperature to 1373 K in 10 hours, maintained at this temperature for 25 hours and slowly cooled down to 773 K in 72 hours. After another 24 hours the ampoule is centrifugalized. The obtained crystals after centrifugation were treated with a 1:1 HCl/H$_2$O solution for 5 hours to remove the redundant tin flux. The single crystal X-ray diffraction was performed with Cu Kα radiation at room temperature using a Rigaku Miniex diffractometer.

*ARPES measurements*

The soft X-ray ARPES measurements on *m*-NiP$_2$ were performed at Beamline 25SU of the SPring-8 synchrotron light source [31], which is equipped a Scienta DA30 electron analyzer. The overall energy and angular resolution were set to be better than 80 meV and 0.2°, respectively. The sample is cleaved *in-situ* by top-post method along the (100) plane at 80 K. During the measurement, the temperature of the sample was kept at 80 K, and the pressure was maintained at less than 2.5 × 10$^{-8}$ Pa. The incident beam is left-circular polarized.

VUV ARPES measurements on *m*-NiP$_2$ were performed at Beamline 13U of the National Synchrotron Radiation Laboratory (NSRL) and Beamline 03U of the Shanghai Synchrotron Radiation Facility (SSRF) [32, 33], both equipped a Scienta DA30 electron analyzer. The overall energy and angular resolution was set to be better than 30 meV and 0.2°, respectively. For experiments at NSRL, the samples are cleaved *in-situ* by top-post method along the (100) plane. The base temperature was set to be 8 K and the base pressure was better than 6 × 10$^{-11}$ torr. For experiments at SSRF, the samples are cleaved *in-situ* by top-post method along both the (11-1) and (100) planes. The temperature was set to be 26 K for samples cleaved along the (11-1) plane and 35 K for those cleaved along the (100) plane. The base pressure was better than $6 \times 10^{-11}$ torr. The temperature-dependent ARPES study was performed at NSRL from $T = 8$ K to 150 K, and the pressure was maintained at less than $1.5 \times 10^{-10}$ torr. The incident beams at the two beamlines are both *p*-polarized.

*First-principles calculations*

The electronic structure calculations were carried out by the DFT method encoded in the Vienna Ab-initio Simulation Package (VASP) [34, 35] based on the projector augmented wave (PAW) method [36]. The Perdew-Burke-Ernzerhof (PBE) approximation is used for the exchange-correlation function [37]. The plane-wave cutoff energy was set to be 520 eV and the DFT-D3 method was also included for the van-der-Waals correction [38]. GGA+U correction was applied to the Ni 3*d* orbitals, and U was set to be 3.0 eV. The *k*-point sampling is $11 \times 11 \times 6$ with the Γ scheme for bulk structure. To study the surface states of the crystal, we constructed a slab structure with the thickness of 8-unit cells. The in-plane lattice constants were set to be $a = 5.629$ Å and $b = 5.616$ Å. The whole Brillouin-zone was sampled by a $6 \times 6 \times 1$ Monkhorst-Pack grid. The experimental values are used for cell parameters [27]. Atomic positions are fully relaxed until the force on each atom was smaller than $1\times10^{-3}$ eV/ Å, and the total energy convergence criterion was set to be $1\times10^{-7}$ eV.

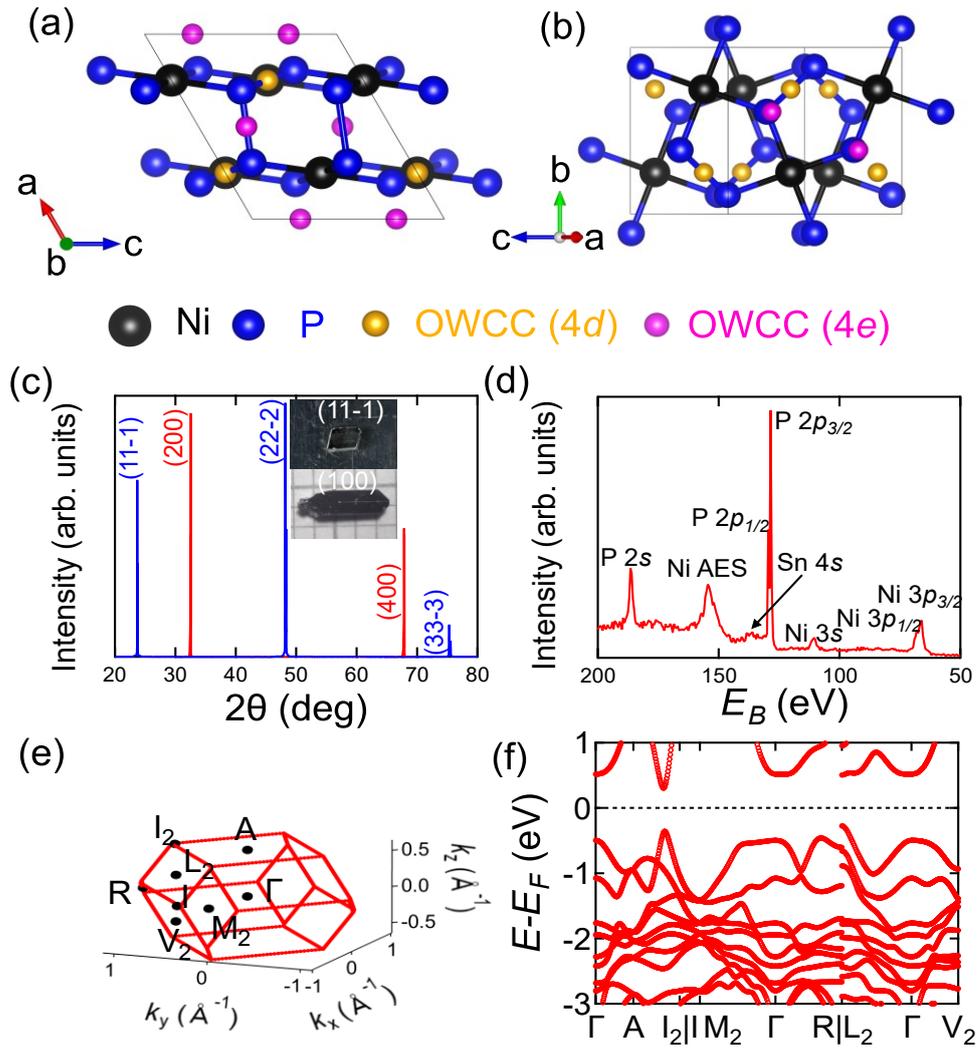

**Fig. 1 Basic properties and calculated bulk bands of *m*-NiP$_2$.** (a, b) Illustration of crystal structure of *m*-NiP$_2$ and location of OWCCs along (010) and (100) planes (side view and top view). (c) Single crystal X-ray diffraction results of *m*-NiP$_2$ terminated at the (100) and (11-1) planes. Inset: photographs of single crystals of *m*-NiP$_2$ with different terminations. (d) The photoemission core level measurement results. Core levels of nickel and phosphorous are clearly resolved. The peak at $E_B \sim 154$ eV corresponds to electrons with kinetic energy of about 845 eV, which should be ascribed to be the Auger electron spectroscopy peak of nickel [39]. Another peak at $E_B \sim 137$ eV is the 4*s* peak of tin, which originates from the residual flux on the sample. (e) The three-dimensional Brillouin zone and high symmetry points (black points). (f) The DFT-calculated bulk bands along the high symmetry path defined in Fig. 1 (e).

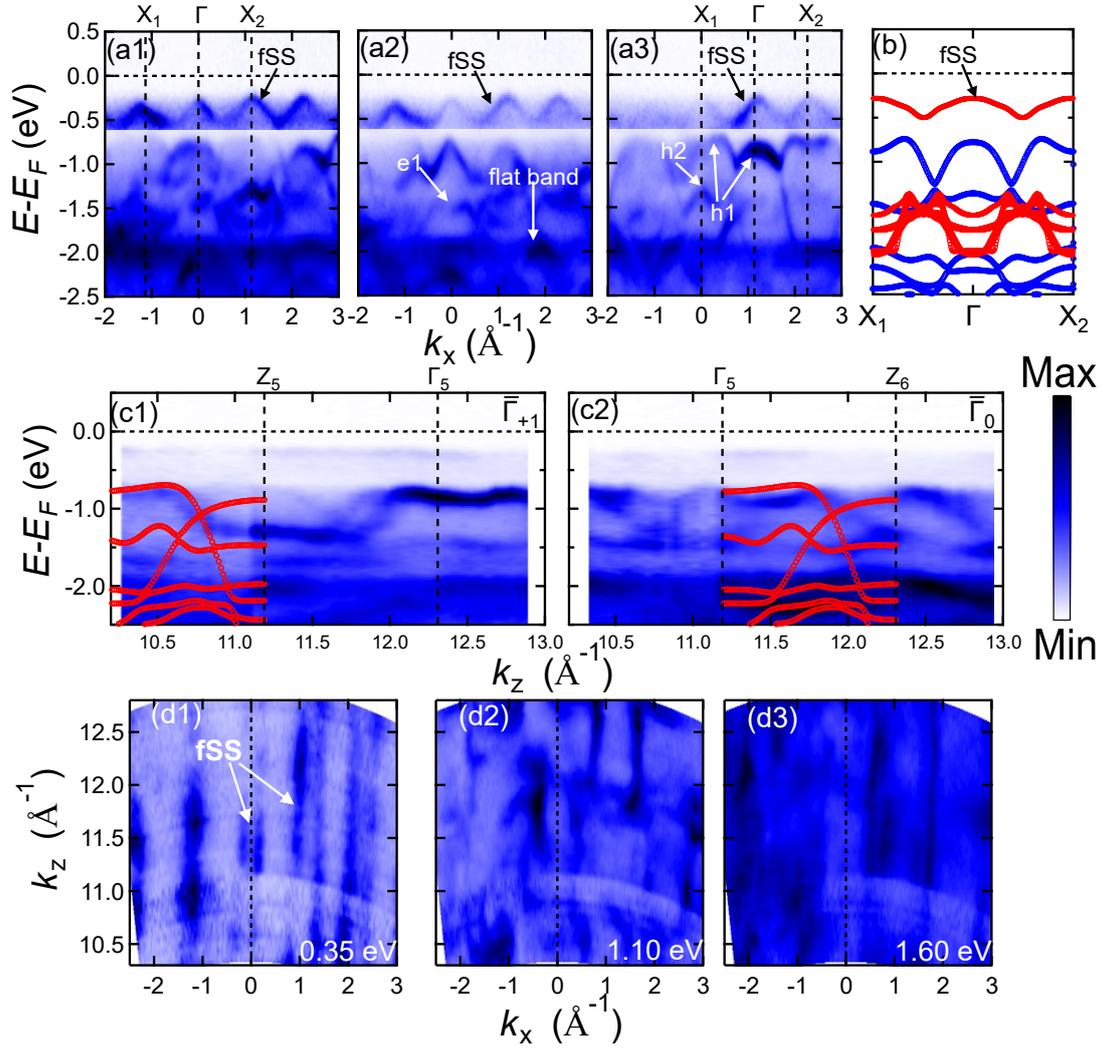

**Fig. 2 The bulk band structure and floating surface state (fSS) of *m*-NiP$_2$ cleaved along the (100) surface.** (a1-a3) The band structure along $\bar{\Gamma}$-X at $k_z$ = 11.2 (the bulk Γ point), 11.8 and 12.3 Å$^{-1}$ (the bulk Z points), respectively. The data within -0.6 to 0.5 eV is plotted using a narrower color scale to highlight the fSS. (b) The DFT calculation results for bulk states (blue dots) and the fSS (red dots) along the $k_x$ direction. (c1, c2) The $E_B$-$k_z$ dispersion collected at the $\bar{\Gamma}$ point of the second ($\bar{\Gamma}_{+1}$, $k_x$ = 1.119 Å$^{-1}$) and the first ($\bar{\Gamma}_0$, $k_x$ = 0) sBZ. The red dots represent the DFT calculation results for the bulk bands. (d1-d3) The constant energy contours in the $k_z$-$k_x$ map at $E_B$ = 0.35, 1.10 and 1.6 eV, respectively. From Fig. 2, we see that the (100)-cleaved *m*-NiP$_2$ processes a surface state that is completely separated from the bulk bands at $E_B$ ~ 0.3 eV, together with a flat bulk band at $E_B$ ~ 2 eV.

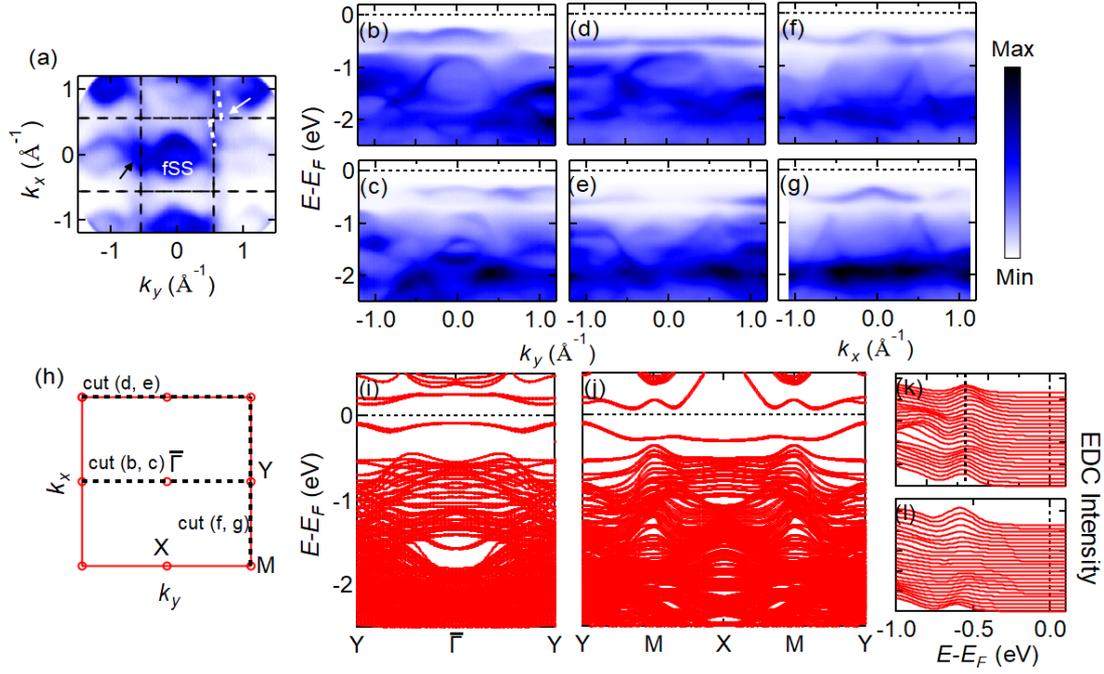

**Fig. 3 In-plane dispersion of the fSS.** (a) ARPES constant energy contour at $E_B$ = 0.4 eV of $m$-NiP$_2$ collected at $h\nu$ = 116 eV ($k_z$ = 5.694 Å$^{-1}$). The black dashed lines indicate the boundary of the sBZ. (b-g) ARPES band maps along in-plane directions marked in Fig. 3(h), collected with (b,d,f) 116 eV ($k_z$ = 5.694 Å$^{-1}$) and (c,e,g) 83 eV ($k_z$ = 4.875 Å$^{-1}$) photons. (h) Schematic illustration of the sBZ on the (100) plane of $m$-NiP$_2$, along with the high symmetric points. (i, j) DFT calculation results along the high symmetric paths defined in Fig. 3 (h), based on an 8-unit-cell slab model. (k, l) The EDCs corresponding to the spectra shown in Fig. 3 (d, e), respectively. The curves are collected from $k_y$ = -1.2 to 1.2 Å$^{-1}$, each integrated within a momentum range of 0.04 Å$^{-1}$. From Fig. 3, it is seen that the fSS of $m$-NiP$_2$ has a narrow global bandwidth of ~300 meV, which becomes even flatter at the edges of the sBZ.

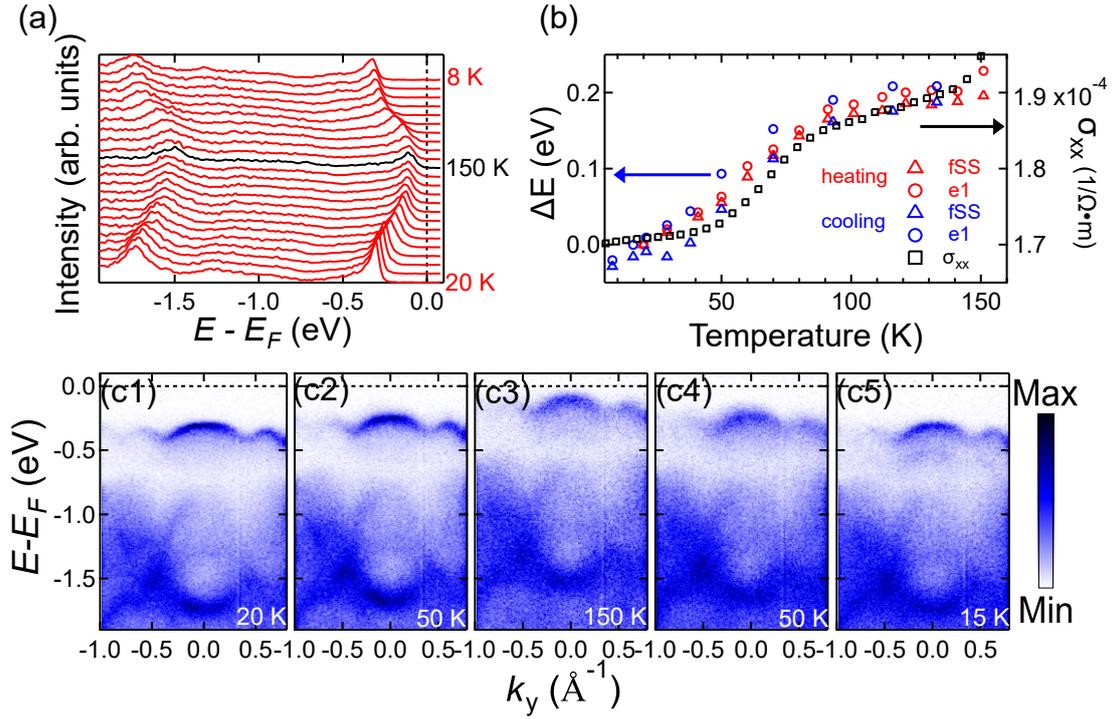

**Fig. 4 Temperature-dependence of energy bands and its correspondence with the electrical conductivity.** ARPES data is collected at $h\nu = 42$ eV. (a) EDCs at different temperatures collected within 0.05 Å$^{-1}$ range near the $\bar{\Gamma}$ point. The temperature cycle starts from 20 K, slowly heats up to 150 K and then cools down back to 8 K. Similar overall shape of the EDCs suggests a rigid-band shift scenario. (b) The band shift of the fSS and the e1 pockets in the heating-cooling temperature cycle (red and blue markers, left axis). The band shifts are obtained by fitting the EDC peaks at different temperatures using Lorentzian functions. The temperature dependence of electrical conductance (black markers, right axis) from $T = 5$ to 150 K is also plotted in the figure, which agrees with the band shifts. (c1-c5) The ARPES $E_B$-$k_y$ maps collected at 20 K (heating), 50 K (heating), 150 K (heating), 50 K (cooling) and 15 K (cooling), respectively. The common trend of temperature evolution between the conductivity of the system and the rigid shifts of the bands shown in Fig. 4 indicates a close relationship between the transport behavior and the electronic structure, especially the fSS which evolves closer to $E_F$ at ambient temperatures.

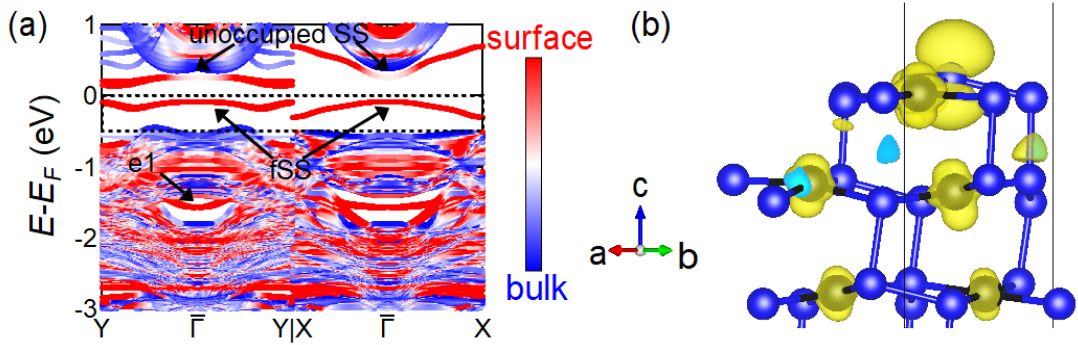

**Fig. 5 Band dispersion and charge distribution of the fSS obtained by slab calculations.** (a) The fully relaxed, surface projected DFT-calculated band structure on an 8-unit-cell slab model. The fSS hole pockets at the $\bar{\Gamma}$ point and the Y point, and the e1 pocket at $E_B \sim 1.5$ eV, are identified as surface states. (b) The DFT-calculated electrical charge distribution (yellow bubbles) of the bands within $E_B = 0$ to 0.5 eV (marked by dashed rectangles in Fig. 5 (a)) on an 8-unit-cell slab model. The electrical charge mainly locates either around the Ni atoms or above the P atoms at the interface (correspond to OWCCs at 4$e$). From Fig. 5, it is inferred that the fSS, together with an additional, unoccupied surface state revealed above $E_F$ (Panel a), correspond to the occupied and unoccupied branches of the OWCC-induced obstructed surface states that are split due to surface reconstruction.